\documentclass[fleqn,twoside]{article}

\usepackage{espcrc2}

\newcommand{\be}{\begin{equation}}
\newcommand{\ee}{\end{equation}}
\newcommand{\e}{{\rm e}}
\newcommand{\Gtbar}{{\overline{G(t)}}}
\newcommand{\Gbar}{{\overline{G}}}
\newcommand{\Gth}{{G_{\rm th}}}
\newcommand{\tp}{{t^\prime}}

\newcommand{\tmin}{{t_{\rm min}}}
\newcommand{\Fig}[1]{Fig.~\ref{#1}}
\newcommand{\eq}[1]{Eq.\,(\ref{#1})}
\newcommand{\chia}{{\chi^2_{\rm aug}}}
\newcommand{\chip}{{\chi^2_{\rm prior}}}

\newcommand{\alphaP}{\alpha_P}
\newcommand{\wbar}{{\overline w}}

\title{Constrained Curve Fitting}

\author{G.\ P.\ Lepage\address{Physics Department, Cornell University, Ithaca, NY, USA}, B.\ Clark\address[CMU]{Physics Department, Carnegie Mellon University, Pittsburgh, PA, USA}, 
C.\ T.\ H.\ Davies\address{Physics Department, University of Glasgow, Glasgow, UK},
K.\ Hornbostel\address{Physics Department, Southern Methodist University, Dallas, TX, USA},
P.\ B.\ Mackenzie\address{Theory Group, Fermi National Accelerator Center, Batavia, IL, USA},
C.\ Morningstar\addressmark[CMU],\\
H.\ Trottier\address{Physics Department, Simon Fraser University, Vancouver, BC, Canada}
}

\begin{document}

\begin{abstract}
We survey techniques for constrained curve fitting, based upon Bayesian statistics, that offer significant advantages over conventional techniques used by lattice field theorists.
\end{abstract}
\maketitle

\section{INTRODUCTION}

With recent developments in lattice QCD, a wide range of high-precision calculations is now possible. To achieve high precision we need tight control of all the systematic errors inherent in the analysis of simulation data. A particularly irritating source of such systematic errors has been curve fitting, which plays a central role in almost all applications of lattice simulations.

Lattice theorists use curve fitting in many different ways: we use it, for example, to extract hadronic masses and matrix elements from
Monte Carlo estimates of correlators, to extrapolate light-quark masses
to physical values, and to fit short-distance Monte Carlo results to
perturbative expansions. In each such case the theory we fit to Monte Carlo data has an infinite number of parameters. In order to fit such a theory using a finite amount of data, we normally truncate both the data set and the theory. Thus we might retain only data for the smallest quark masses, and extrapolate linearly to physical masses. Or we might fit only the large-$t$ behavior of a hadronic correlator, ignoring contributions from all but the lowest-mass hadron.
Such truncations are usually necessary to obtain good fits with reasonable
error estimates for the fit parameters, but they 
necessarily increase both the statistical and systematic uncertainties
in the results. Statistical uncertainties are increased because Monte
Carlo data is discarded. Systematic uncertainties are introduced by
omitting parts of the theory that could conceivably be significant. Such
systematic effects have proven particularly difficult to estimate in
past analyses, and these are the principle focus of this article.

In this paper we show how to circumvent the shortcomings of the traditional approach by using constrained curve fitting, a simple modification of standard
maximum likelihood techniques\,\cite{othertalks}. Constrained curve fits, while widely used in other research areas, are not used much by the lattice QCD
community. Here we will show how they can be adapted for use in lattice
calculations, and we will document their striking advantages over
traditional methods. Constrained
curve fits provide an elegant procedure for incorporating
systematic uncertainties due to underconstrained parts of a theory
(high-energy states, for example). Furthermore they allow us to fit much more data\,---\,for example, correlators down to $t=0$. And they are numerically more robust than unconstrained fits. Finally the formalism for constrained curve fitting can be used to estimate the lattice spacings and quark masses that will minimize the errors in a large-scale simulation.

\section{CONSTRAINED FITS}

\subsection{The Problem}
The central problem is illustrated by the analysis of a meson correlator. Simulations generate a Monte Carlo estimate, $\Gtbar$, of the correlator for a finite number of time steps, say $t=0,1\ldots23$.  Theory tells us that the exact correlator has the form
\be \label{gth}
\Gth(t;A_n,E_n) = \sum_{n=1}^{\infty} A_n\,\e^{-E_nt},
\ee
where we assume that the energies~$E_n$ are in order of increasing size.
The challenge is to fit an infinite number of amplitudes~$A_n$ and energies~$E_n$ using only 24~$\Gtbar$'s.

Traditional fits minimize $\chi^2(A_n,E_n)$ by varying~$A_n$ and~$E_n$, where
\be \label{chi2}
\chi^2(A_n,E_n) \equiv \sum_{t,\tp} \Delta G(t) \, \,\sigma^{-2}_{t,t^\prime}\, 
\Delta G(\tp),
\ee
and
\be
\Delta G(t) \equiv \Gtbar - \Gth(t;A_n,E_n).
\ee
The correlation matrix is estimated from the Monte Carlo:
\be
\sigma^2_{t,t^\prime} \,\equiv\, \overline{G(t)G(t^\prime)} - 
\overline{G(t)}\,\,\overline{G(t^\prime)}.
\ee
Unfortunately this fitting procedure is singular here since there are more fit parameters, $A_n$ and $E_n$, than data; the final uncertainties in the fit parameters are infinite.  Additional information is needed if we are to proceed.

The information we normally add is that the $A_n$'s are well behaved, and therefore contributions from high-energy states are suppressed at large~$t$ by the exponentials in the correlator, 
\eq{gth}. Thus there is a~$\tmin$ above which only the first one or two terms in $\Gth$ make statistically significant contributions. 
The standard procedure therefore is to retain, say, only the first two terms in 
$\Gth$ and to fit them using only Monte Carlo results from $t\ge\tmin$. The trick is to find the best $\tmin$. Choosing $\tmin$ too small biases the $E_n$'s and $A_n$'s away from their true values, introducing systematic errors (because two terms is not enough in~$\Gth$). Choosing $\tmin$ too large gives statistical errors~$\sigma_{E_n}$ and~$\sigma_{A_n}$ that are too large, since useful data is discarded. One typically tries to increase $\tmin$ until the statistical errors mask any possible systematic error. Without a reliable quantitative estimate of the systematic error, however, any procedure for setting $\tmin$ is necessarily {\it ad hoc}.

\begin{figure}
\input{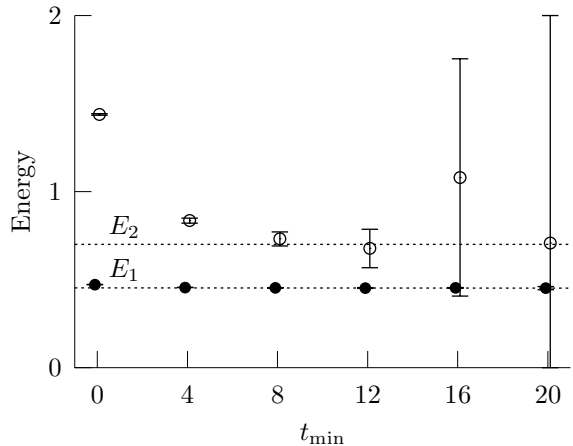}
\caption{Fit values for the lowest two energies from a 2-term fit to a local-local $\Upsilon$ correlator using different $\tmin$'s. The correct values, from other analyses, are indicated by the dotted lines.
}
\label{fig:tmin}
\end{figure}

To illustrate the dependence on $\tmin$, we plot results for 
$E_1$ and $E_2$ from 2-term fits with various $\tmin$'s in 
\Fig{fig:tmin}. The Monte Carlo data for these fits was obtained by averaging 840 $\Upsilon$ correlators evaluated at (quenched) $\beta=6$ with local sources and sinks\,\cite{davies}. The competition between large systematic errors for small $\tmin$ and large statistical errors for large $\tmin$ is particularly apparent for~$E_2$ in this plot.

\subsection{A Solution}
\label{sec:constrained}
\begin{figure}
\input{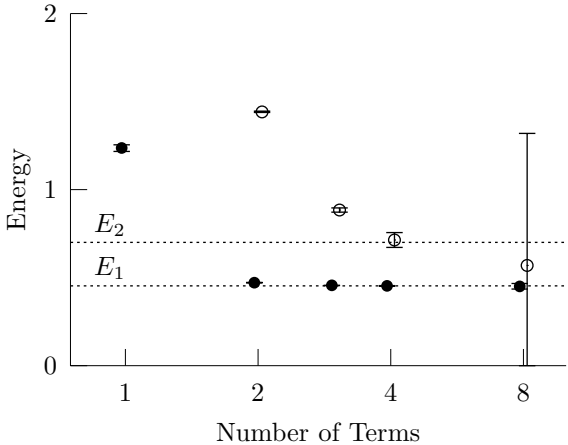}
\caption{Fit values for the two lowest energies from unconstrained fits with different numbers of terms in $\Gth$. The correlator is a local-local $\Upsilon$ correlator and is fit for all $t$'s.
}
\label{fig:mmax}
\end{figure}

Our goal should be to fit all the Monte Carlo data ($\tmin=0$) using as many terms as we wish in $\Gth$. As we add more terms to $\Gth$, however, the errors on the leading parameters grow steadily in a traditional analysis, as is evident in \Fig{fig:mmax}. The reason is easily understood. The large uncertainties in $E_1$ and $E_2$ for the 8-term fit, for example, result because the parameters for higher-energy states are poorly constrained by the data and therefore wander off to unphysical values. 
Thus amplitude $A_4$ ranges between five and ten times~$A_1$ in the 8-term fit, while quark models suggest that $A_4$ is of order $A_1$ or smaller. Since the allowed range for $A_4$ affects the error estimates for other parameters, the errors on $E_1$ and $E_2$ will be unreasonably large so long as the fitting code assumes that $A_4\approx 10A_1$ is plausible.  We need some way to teach physics to the fitting code.

To constrain fit parameters to physically reasonable ranges, we augment 
the $\chi^2$ before minimizing:
\be
\chi^2 \quad\to\quad \chia \equiv \chi^2 + \chip,
\ee
where
\be \label{chip}
\chip \equiv \sum_n \frac{(A_n-\tilde{A}_n)^2}{\tilde{\sigma}^2_{A_n}}
+ \sum_n \frac{(E_n-\tilde{E}_n)^2}{\tilde{\sigma}^2_{E_n}}.
\ee
The extra terms in $\chia$ favor $A_n$'s in the interval $\tilde{A}_n\pm
\tilde{\sigma}_{A_n}$ and $E_n$'s in $\tilde{E}_n\pm\tilde{\sigma}_{E_n}$. The 
$\tilde{A}_n$'s, $\tilde{\sigma}_{A_n}$'s\,\ldots\,are {\em inputs\/} to the fitting procedure. We choose reasonable values for them on the basis of 
{\em prior\/} knowledge. This set of input parameters is referred to collectively as the ``priors.''

\begin{figure}
\input{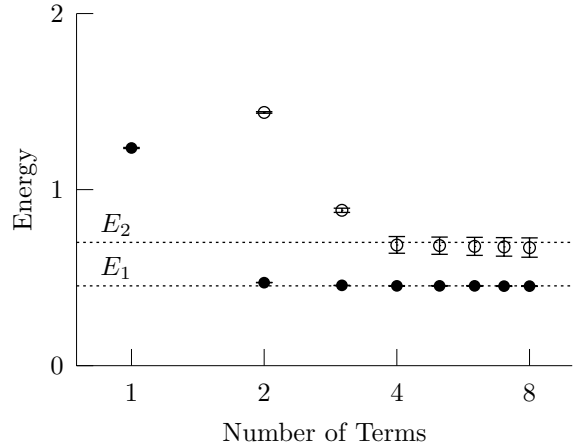}
\caption{Fit values for the two lowest energies from constrained fits with different numbers of terms in $\Gth$. The correlator is a local-local $\Upsilon$ correlator and is fit for all $t$'s.
}
\label{fig:constrained}
\end{figure}

Having chosen the priors, the procedure for a constrained fit is to 
minimize~$\chia$ fitting all of the Monte Carlo data ($\tmin=0$). 
The number of terms in 
$\Gth$ is increased until fit results converge for the parameters of interest. Unlike~$\tmin$, the number of terms in~$\Gth$ need not be optimized; it is simply increased until results converge. This is illustrated by fit results for $E_1$ and 
$E_2$ from our $\Upsilon$~data, which are plotted in \Fig{fig:constrained} for fits with different numbers of terms.

The numerics are greatly improved by the constraints. For example, one can easily fit 100 terms in $\Gth$ to the $\Upsilon$~data, even though there are only 24 data points. The fit results for all but the first few parameters simply reproduce the prior information in such a highly overparameterized fit.

The error estimates for the fit parameters in our $\Upsilon$~fits automatically combine both the statistical errors in the Monte Carlo data, and the systematic errors due to our limited knowledge, as specified by the priors, about the poorly constrained high-energy states.  To see how much error is due to each source we refit with 
all $\tilde{\sigma}$'s doubled. Results for the energies change as follows:
\be
\begin{array}{lcl}
E_1 = 0.4526 (15) & \to& 0.4528 (14) \\
E_2 = 0.683 (49)  & \to& 0.697 (50) \\
E_3 = 1.05 (12) &\to &1.10 (21).\\
\end{array}
\ee
Evidently $E_1$ and $E_2$ are determined largely by the Monte Carlo data, while $E_3$ is strongly affected by the priors. The insensitivity of the leading parameters to the priors is typical for high-quality data. The result for $E_2$ is impressive given that it comes from a single local-local correlator. (It also agrees with results obtained from multi-source/sink fits\,\cite{davies}.)

We actually parameterized our $\Upsilon$~fits in terms of parameters $a_n\equiv\log A_n$ and $\varepsilon_n\equiv\log(E_n-E_{n-1})$. This parameterization builds in 
{\em a priori} requirements, $A_n\!>\!0$ and $E_n\!>\!E_{n-1}$, which improve the fits. Using previous simulations as a guide, we chose priors that favored 
$a_n\approx \log0.02\pm\log2$ and $\varepsilon_n \approx \log0.2 \pm \log2$ or:
\be
A_n \approx 0.02 \begin{array}{c} +0.02 \\ -0.01 \end{array}
\quad
E_n-E_{n-1} \approx 0.2 \begin{array}{c} +0.2 \\ -0.1 \end{array}.
\ee

\begin{figure}
\input{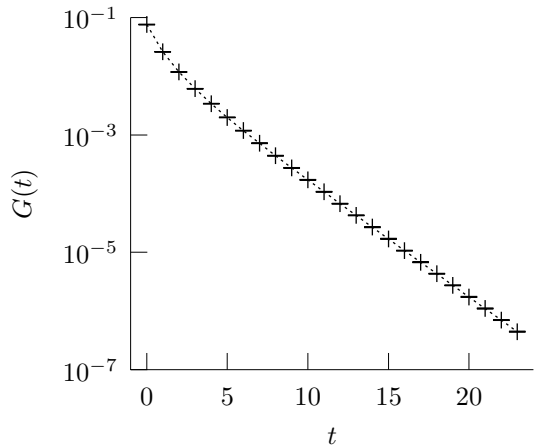}
\caption{The constrained 5-term fit to the local-local $\Upsilon$ correlator. The statistical errors in the data points are too small to be resolved in the plot.}
\label{fig:ups}
\end{figure}

Our best 5-term fit to the $\Upsilon$~data is shown in \Fig{fig:ups}, together with the data. The fit is excellent all the way down to $t=0$: the minimum $\chia$ divided by the number of data points is~0.8. (This is the correct ratio to examine since $\chia$ has one extra term beyond those in $\chi^2$ for each fit parameter.) While $\chi^2$, $Q$, {\em etc.} play a crucial role in traditional fits, where $\tmin$ must be optimized, they are of secondary importance in constrained curve fitting. The key criterion is convergence as the number of parameters is increased. If $\chia$ per data point is significantly larger than~1 after convergence, then there is likely a mistake in either the data or the theory. 

\begin{figure}
\input{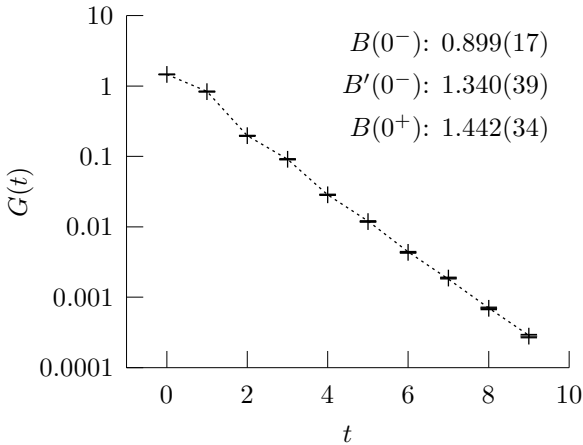}
\caption{A constrained fit to a local-local $B$ meson correlator made with NRQCD and staggered quark propagators. The energies of the three lowest-energy states are listed in lattice units. The statistical errors in the data points are too small to be resolved in the plot.
}
\label{fig:staggB}
\end{figure}

We have experimented with constrained fits for a wide variety of other correlators, including fits of 30--40 parameters for $4\times3$ matrix $G$'s, simultaneous fits of multiple channels ({\em e.g.}, $\pi$ and $\rho$), static potentials and glueball masses, and correlators involving staggered quarks. All work well. In some cases fits are greatly simplified. An example is the fit shown in \Fig{fig:staggB} for a $B$~meson correlator made from an NRQCD propagator for the $b$~quark and a staggered quark propagator for the $d$~quark\,\cite{shigemitsu}. Such fits are complicated by contributions from opposite parity states, introduced by the staggered quarks,  that oscillate with an overall factor $(-1)^t$. Traditional fits have difficulty quantifying the importance of these contributions since they are small at 
large~$t$. Constrained fitting down to $t=0$, however, makes it easy to account and correct for them.

\section{THEORY AND INTERPRETATION}

\subsection{Bayesian Statistics}
Bayesian statistics provides a useful framework for understanding the assumptions that go into constrained curve fitting\,\cite{sivia}. In this approach the analysis is recast in terms of probabilities: $P(A|B)$ denotes the probability that $A$ is true or correct assuming $B$ is true. 
For compactness we denote the set of all parameters by $\rho = \{A_1,E_1\ldots\}$, the set of prior parameters by $\eta
=\{\tilde{A}_1,\tilde\sigma_{A_1}\ldots\}$, and the Monte Carlo data by~$\Gbar$.

We begin with two assumptions. The first is that the Monte Carlo data set is sufficiently large that $\Gbar$~has Gaussian statistics (Central Limit Theorem). Then the probability density for obtaining a particular $\Gbar$ given a particular theory, specified by $\rho$, is
\be
P( \Gbar | \rho) \propto \e^{-\chi^2(\rho)/2}
\ee
where $\chi^2$ is defined as in~\eq{chi2}.\footnote{This is not strictly correct since the correlation matrix $\sigma_{t,\tp}^2$ in $\chi^2$ will in general depend upon $\rho$, while in practice we use a fixed Monte Carlo estimate of it.} 

What we need ultimately is the probability that a particular set of parameters $\rho$ is correct given the Monte Carlo data\,---\,that is, we need $P(\rho|\Gbar)$, not $P(\Gbar|\rho)$. $P(\rho|\Gbar)$ is connected to $P(\Gbar|\rho)$ by Bayes Theorem:\footnote{This formula follows from the trivial identity for probabilities:
$P(\rho\Gbar) =  P(\Gbar|\rho)\,P(\rho) = P(\rho|\Gbar)P(\Gbar)$.
}
\be \label{bayes}
P(\rho|\Gbar) = \frac{P(\Gbar|\rho)\,P(\rho)}{P(\Gbar)} 
\propto P(\Gbar|\rho)\,P(\rho).
\ee
Here $P(\Gbar)$ is the probability of obtaining a particular $\Gbar$ from any theory; it is $\rho$ independent. More important is $P(\rho)$ which is the probability that a particular set of parameters $\rho$ is correct in the absence of any new data. It contains what we know about the parameters before we begin the fit. This is called the ``prior'' distribution, or simply the prior.

Our second assumption is that the prior distribution can be approximated by the Gaussian
\be
P(\rho) = \e^{-\chip(\rho)/2}
\ee
where $\chip$ is defined as in~\eq{chip}. The {\it a priori\/} assumption therefore is that $\rho_i \approx \tilde{\rho_i} \pm \tilde{\sigma}_{\rho_i}$.  With this assumption our final probability function is
\be \label{finalP}
P(\rho|\Gbar) \propto \e^{-\chia(\rho)/2}
\ee
where $\chia$ is the augmented $\chi^2$ introduced in 
Section~\ref{sec:constrained}.

The choice of a Gaussian for the prior distribution is arbitrary; other choices might well be appropriate. There is, however, an argument that suggests Gaussians. Beyond specifying an average value for each parameter and a standard deviation, we want the prior distribution to be as unbiased as possible. In general the least biased choice for a probability density is the one that minimizes the information content, or, equivalently, maximizes the entropy,
\be
S \equiv -\int P(\rho) \,\log P(\rho) \, d\rho.
\ee
Given the constraints $\langle \rho \rangle = \tilde\rho$ and $\langle \rho^2\rangle -\langle\rho\rangle^2 = \tilde{\sigma}^2$, a simple variational calculation shows that the entropy is maximized by the Gaussian
\be
P(\rho) = \frac{1}{\sqrt{2\pi} \tilde\sigma}\,
\e^{-(\rho-\tilde\rho)^2/2\tilde\sigma^2}.
\ee
This argument is less compelling than it seems, however, because it relies upon the implicit (and arbitrary) assumption that all $\rho$'s are equally likely in the absence of any information about the mean or standard deviation. A different parameterization implies a different assumption.

\subsection{Error Estimation}
Given $P(\rho|\Gbar)$ we can compute everything we want to know using integrals; in principle no minimization is needed. For example, we obtain a statistical estimate of an arbitrary function of the parameters  using
\be \label{bayesintegral}
\langle f(\rho) \rangle = B^{-1} \,\int
\e^{-\chia(\rho)/2}\,f(\rho)\, d^n\rho
\ee
where
\be
B \equiv  \int\e^{-\chia(\rho)/2}\,d^n\rho ,
\ee
and the variance is $\sigma^2_f\equiv\langle f^2 \rangle - \langle f \rangle^2$, as usual. In practice these integrals are quite difficult to evaluate for all but the simplest of fits. This is because $P(\rho|\Gbar)$ is typically very sharply peaked about its maximum. For smaller problems, adaptive Monte Carlo integrators, such as \verb|vegas|, are effective. For larger problems Monte Carlo simulation techniques, such as the Metropolis or hybrid Monte Carlo methods, can be effective.
Still the cost of evaluating the integrals is often prohibitive, particularly when there are lots of poorly constrained parameters (which lead to long, narrow, high ridges in the probability distribution). Consequently efficient approximations are useful.

One approximation assumes that enough statistics have been accumulated so that 
$\chia(\rho)$ is approximately quadratic in the $\rho$'s throughout the dominant integration region:
\be \label{quadratic1}
\chia(\rho) \approx \chi^2_{\rm aug,min} + \sum (\rho-\rho^*)_i(\rho-\rho^*)_j
C^{-1}_{ij}
\ee
where parameters $\rho=\rho^*$ minimize $\chia$.
In this quadratic limit we can approximate
\be \label{quadratic2}
\langle f(\rho)\rangle \approx f(\rho^*),
\:\:
\sigma^2_f \approx \sum C_{ij}\,\partial_i f(\rho^*) \,\partial_j f(\rho^*)
\ee
provided $f(\rho)$ is sufficiently smooth as well.
This is the constrained curve fitting procedure outlined in 
Section~\ref{sec:constrained}.

A different, more robust approximation to the Bayesian integrals uses a bootstrap analysis, suitably modified to account for the prior distribution\,\cite{hornbostel}. A bootstrap analysis generates an ensemble of $\rho$'s whose distribution approximates the Bayesian distribution~$P(\rho|\Gbar)$. Each $\rho$ in this ensemble is obtained by minimizing $\chia$, but with different Monte Carlo data and different means for the priors for each~$\rho$. We replace $\Gbar$ in $\chia$ by the average of a random selection of $G$'s, allowing duplicates, from the original Monte Carlo data ensemble, just as in the standard bootstrap method. The new means, $\tilde\rho^\prime_i$, for the priors are random numbers drawn from a Gaussian distribution  with mean $\tilde\rho_i$ and standard deviation $\tilde\sigma_{\rho_i}$. The new means incorporate the effects of the priors into the bootstrap $\rho$ distribution. Given 100 or 1000 $\rho$'s generated in this fashion, we then average over the ensemble to compute estimates of any function~$f(\rho)$.
Although this procedure entails a minimization for each $\rho$ in the ensemble, we find that is generally much faster than Monte Carlo evaluation of Bayes integrals (\eq{bayesintegral}).

The distribution obtained from this modified bootstrap algorithm is not precisely the Bayes distribution~$P(\rho|\Gbar)$. It has additional factors such as $\sqrt{\det g_{ij}}$ where
\be
g_{ij} \equiv \sum_{t,\tp} \sigma^{-2}_{t,\tp}\, 
\frac{\partial G(t;\rho)}{\partial \rho_i} \,
 \frac{\partial G(\tp;\rho)}{\partial \rho_j}
\ee
is a metric induced on $\rho$~space\,\cite{hornbostel}. These factors become constants for sufficiently high statistics and so make no difference in that limit. This particular factor is interesting, however, because it makes the measure in $\rho$ space invariant under reparameterizations. This suggests that  
\be
P^\prime(\rho|\Gbar) \propto \sqrt{\det g_{ij}}\,\e^{-\chia/2}
\ee
might be a better choice for our Bayesian probability. Then only~$\chip$ would depend upon the parameterization of~$\Gth$, and $\chip$ usually has little influence on the best-determined fit parameters. 

\section{OTHER APPLICATIONS}
\label{sec:pth}
One approach to evaluating high-order QCD perturbation theory is to simulate the quantity of interest at high $\beta$'s, where QCD is perturbative, and fit perturbative expansions to the results. An example, from~\cite{trottier}, is the $5\times5$~Wilson loop, $W(5,5)$, which was measured in Monte Carlo simulations with 9~different lattice spacings that gave couplings $\alpha_P(3.4/a)$ covering the range~0.01--0.07. The measured values of $W(5,5)$ were fit using an expansion
\be \label{w55}
-\mbox{$\frac{1}{20}$}\log W(5,5) = \sum_{n=1}^{M} c_n \alpha_P^n(q^*)
\ee 
where $q^*=2.23/a$ and $M\ge5$. Priors $\tilde{c}_n = 0$ and $\tilde\sigma_{c_n}=5$ were used in the fits to obtain the results in Table~\ref{table:w55} (for $M=5$).
The agreement with exact results is excellent. Refitting with $c_1$ and $c_2$ set equal to their exact values gives $c_3=3.91(21)$, which agrees well with the result in Table~\ref{table:w55} but which is much more precise.
\begin{table}
\begin{center}
\begin{tabular}{llll} \hline
& $c_1$ & $c_2$ & $c_3$  \\ \hline
fit: & 1.7693(6) & $-1.201(40)$ & 4.3(7) \\
exact: & 1.7690 & $-1.177$ \\ \hline
& $c_4$ & $c_5$ \\ \hline
fit: & 0.7(4.9) & 0.1(5.0) \\
exact:  \\ \hline
\end{tabular}
\end{center}
\caption{Perturbative coefficients for $W(5,5)$ as determined from fits to high-$\beta$ simulations, and exactly from Feynman integrals. Refitting with $c_1$ and $c_2$ held at their exact values gives $c_3=3.91(21)$.
}
\label{table:w55}
\end{table}

The errors for the $c_n$'s again include both statistical errors, reflecting the precision of the data, and systematic errors, reflecting the degree of our ignorance of higher-order coefficients. The low-order coefficients ($n\le3$) are largely determined by the Monte Carlo data; their values and errors are essentially unchanged if the $\tilde\sigma_{c_n}$'s are doubled, or if 
$M$~is doubled. The high-order coefficients are controlled mostly by the priors. Were we to increase the statistics, more $c_n$'s would be determined by the data. Thus the effective order of the fit increases {\em automatically\/} as warranted by the statistics. This is one of the most attractive aspects of constrained fitting: We don't have to agonize about whether a fit should be 
linear or quadratic or\ldots; instead we set $M=10$ and allow the data to tell us how much it can determine.

Similar issues arise in extrapolations, for example, to the chiral mass limit or to zero lattice spacing. By using constrained fits, we can escape the limitations of linear or quadratic extrapolations. Instead we can include many orders, with constrained coefficients, and allow the Monte Carlo data to determine the relevant order automatically. (See~\cite{trottier} for an illustration of an extrapolation in lattice size~$L$.) 

\section{VARIATIONS}
\subsection{Optimizing Simulations}

In our numerical fits of perturbation theory to simulation data 
(Section~\ref{sec:pth}),
the final errors for the various perturbative coefficients, $c_n$, are affected by the $\alphaP$'s at which we chose to simulate. Our fitting formalism can be used to determine the choice of $\alphaP$'s that minimizes these errors, thereby optimizing the simulation\,\cite{trottier}.  
To illustrate how this is done, we consider a simpler problem. We measure a quantity~$w$ whose perturbative expansion is 
\be
w = 1+c_1\alphaP+c_2\alphaP^2+\cdots.
\ee
For simplicity we ignore all terms beyond second order, and we limit ourselves to a single measurement of~$w$. The challenge is to determine which value of~$\alphaP$ (and therefore which lattice spacing) will lead to the smallest errors when we determine $c_1$ from a Monte Carlo measurement, $\wbar\pm\sigma_w$, of~$w$. This error can be determined from the second derivatives of 
\be
\chia \equiv \frac{(1+c_1\alphaP +c_2\alphaP^2 - \overline w)^2}{\sigma_w^2}
+\frac{c_1^2 + c_2^2}{\tilde\sigma_c^2}
\ee
using Eqns.\,(\ref{quadratic1}) and~(\ref{quadratic2}).   The last term in this equation is our prior; we expect $c_n\approx 0 \pm \tilde\sigma_c$.
We obtain the following formula for the fit error on $c_1$:
\be
\sigma_{c_1}^2  = \frac{
	(\alphaP^4\tilde\sigma_c^2+\sigma_w^2)\tilde\sigma_c^2
}{ (\alphaP^4 + \alphaP^2)\tilde\sigma_c^2 + \sigma_w^2 },
\ee
where we will assume that $\sigma_w$~is independent of~$\alphaP$ (which is {\em not\/} true for Wilson loops, for example, unless numerical roundoff errors dominate). Note that $\sigma_{c_1}\to\tilde\sigma_c$ in either of the limits $\alphaP\to0$ or $\alphaP\to\infty$;  we  would learn nothing new from a simulation in either limit.
The error is minimum at
\be
\alpha = \alpha^* \equiv \sqrt{\sigma_w/\tilde\sigma_c},
\ee
where 
$\partial\sigma_{c_1}/\partial\alpha$ vanishes. At this value
\be
\sigma^2_{c_1} \to \tilde\sigma_c^2\,\frac{2\sigma_w}{\tilde\sigma_c+2\sigma_w}.
\ee
Thus $\alpha=\alpha^*$ is the preferred value for a simulation; the optimal value depends upon the amount of computing that is available (to reduce $\sigma_w$).

The $c_1$ error decreases like $\sqrt{\sigma_w}$ as the simulation 
error,~$\sigma_w$, vanishes. This is rather slow; we would prefer an error that vanishes like $\sigma_w$. At some small value of $\sigma_w$ it will be better to divide the available computer resources between simulations at two values of~$\alphaP$, rather than just one. The analysis above can be repeated for this case, to determine the two optimal $\alphaP$'s, and the resulting $\sigma_{c_1}$ can be compared with that above, to determine when to switch from one~$\alphaP$ to two.

Our simple analysis illustrates a generic design step that should precede any large-scale simulation in lattice QCD. By examining $\chia$'s for important quantities, we can estimate the optimal lattice spacings and quark masses that should be used in the simulation 
{\em before\/} we simulate. 
Furthermore we can estimate the optimal fraction of the total computing resource to spend on each case.  Simulation costs  should be significantly reduced by such optimizations. 

\subsection{Tuning Priors: Empirical Bayes}
The denominator in the Bayes expression, \eq{bayes}, is the probability of obtaining the Monte Carlo average $\Gbar$ given the prior information:
\be
P(\Gbar) = P(\Gbar|\eta) \propto \int 
\frac{\e^{-\chia/2}}{\prod\tilde\sigma_{\rho_i}} \,d^n\rho,
\ee
which, in the Gaussian limit (\eq{quadratic1}), becomes
\be \label{gbf}
P(\Gbar|\eta) \propto  \frac{\sqrt{\det C_{ij}}}{\prod \tilde\sigma_{\rho_i}}
\,\,\e^{-\chi^2_{\rm aug,min}/2}.
\ee
This probability will be small if the priors are inconsistent with the data. This suggests that one might tune the prior parameters~$\eta_i$ to maximize probability~$P(\Gbar|\eta)$, thereby determining the priors from the data. This is referred to as the ``empirical Bayes'' method. It can be useful if only one or two prior parameters is being optimized. For example, when fitting the perturbation series in~\eq{w55}, one might vary all the $\tilde\sigma_{c_n}$'s together to find the $\tilde\sigma_c$ that maximizes $P(\Gth|\eta)$. (Note that the Gaussian approximation, \eq{gbf}, is exact in this case).  Then one would fit with all 
$\tilde\sigma_{c_n}=\tilde\sigma_c$. Note, however, that separately optimizing all 
$\eta_i$'s leads to complete nonsense.\footnote{The ``best'' fit is then the $\rho$ that minimizes $\chi^2$ (not $\chia$), and the fit errors are zero!}

\subsection{Nonparameteric Fits and Maximum Entropy}
Bayesian methods have previously been used for lattice data, and indeed for fitting correlators\,\cite{mem}. The priors and parameterizations used in these earlier papers, however, are quite different from what we have discussed above. These papers express the correlator in terms of its spectral function,~$\rho(\omega)$:
\be
G(t) \equiv \int d\omega \,\e^{-\omega t} \,\rho(\omega).
\ee
The $\omega$ axis is divided into a large number of increments, say 750 or 1000, with centers~$\omega_j$, and the integral is approximated by a sum. The fit parameters are the values of $\rho(\omega_j)$ for all~$j$. This is an example of a ``nonparametric'' fit. Peaks in $\rho(\omega)$ signal bound states. Since the number of fit parameters is typically (much) larger than the number data points, a prior distribution is essential. 

The Maximum Entropy principle provides a framework for constructing priors for this purpose. The spectral function shares several properties with probability densities: it has random noise, $\rho(\omega)\ge0$, and $\rho(\omega) d\omega$ is physically meaningful. This suggests that we think of the spectral function as a probability density. The most probable density is the one with the largest entropy, which suggests that we choose a monotonic function of the entropy as a prior: for example,
\be
P(\rho) \propto \e^{\alpha\,S(\rho)}
\ee
is maximum when the entropy~$S(\rho)$ is maximum.  Using a prior of this form we can fit all of the $\rho(\omega_j)$'s to the Monte Carlo data. (Parameter
 $\alpha$ can be optimized using the empirical Bayes method; see the previous section.)
 
 This algorithm has been used to fit the same $\Upsilon$ data that we used for our constrained fits\,\cite{oevers}. The errors that result for the first excited state ($E_2$) are 
 2--3~times larger than we obtained from our constrained fits in Section~\ref{sec:constrained}. This is not surprising. The maximum entropy prior contains far less information than we included in the priors for the constrained fits, and that is its weakness in this context. One should use all the prior information one has when fitting data. The maximum entropy prior is useful when there is little initial information and no simple parameterization of the theory, as is the case in image processing, for example.

\section{CONCLUSION}
Constrained curve fitting is a major improvement over the traditional fitting procedures used in lattice QCD analyses. Neither the data nor the theory is truncated. Consequently more information, about excited states or higher-order terms, is extracted from the same Monte Carlo simulations. Most importantly these techniques provide a transparent and controlled procedure for including systematic errors due to the infinitely many parameters in the theory that are not constrained by the Monte Carlo data. Constrained curve fitting is useful for any kind of analysis that involves correlators. It is also useful for extrapolating to small quark masses or lattice spacings; the order of extrapolation increases automatically to match the statistical precision of the data. Finally we can use the formalism to estimate the quark masses and lattice spacings that optimize a large-scale simulation given a finite computing resource.

Priors make some people uneasy because the constraints are added
information, beyond the Monte Carlo data, that can bias the fit
results. Such people might argue that one should ``let the data speak
for itself.''  However, it is precisely our ability to fold extra
information into the fit that makes constrained curve fitting so much
more powerful and useful than unconstrained fitting. 
Better {\it a priori\/} knowledge, for example from earlier analyses, results in smaller $\tilde\sigma$'s and smaller systematic errors. No 
{\it a priori\/} knowledge, which corresponds to the unconstrained case,
(correctly) results in infinite systematic errors. Traditional fitting is a special case of constrained fitting, with infinitely broad priors for the first few parameters and infinitely narrow priors, centered around zero, for the others. The priors we use in our constrained fits are far more reasonable. 

\section*{ACKNOWLEDGMENTS}
We thank Persis Drell, Saul Teukolsky, and, especially, Tom Loredo for introducing us to Bayesian techniques. 
We thank Junko Shigemitsu and Matthew Wingate for the use of their Monte Carlo data and for many useful discussions. 
This work was supported in part by the National Science Foundation, and by the Department of Energy.


\begin{thebibliography}{99}
\bibitem{othertalks} Constrained curve fitting is also discussed in the talks at this meeting by Christine Davies, Martin G\"{u}rtler, and Matthew Wingate.
\end{thebibliography}
\end{document}